\documentclass[
superscriptaddress,
twocolumn,
amsmath,amssymb,
longbibliography,
aps,
prl,
]{revtex4-1}
\usepackage{graphicx}
\usepackage{dcolumn}
\usepackage{bm}
\usepackage{color}
\usepackage{hyperref}
\usepackage{xcolor}

\newcommand{\ef}[0]{\mathcal{E}_{\mathrm{F}}}
\newcommand{\teff}[0]{\tau_{\mathrm{eff}}}

\begin{document}


\title{Propagation-induced limits to high harmonic generation in 3D Dirac semimetals}

\author{Jeremy Lim}
\affiliation{%
Science, Math and Technology, Singapore University of Technology and Design, 8 Somapah Road, Singapore 487372, Singapore
}%

\author{Yee Sin Ang}
\affiliation{%
Science, Math and Technology, Singapore University of Technology and Design, 8 Somapah Road, Singapore 487372, Singapore
}%

\author{Lay Kee Ang}
\affiliation{%
Science, Math and Technology, Singapore University of Technology and Design, 8 Somapah Road, Singapore 487372, Singapore
}%

\author{Liang Jie Wong}
\email{liangjie.wong@ntu.edu.sg}
\affiliation{
School of Electrical and Electronic Engineering, Nanyang Technological University, 50 Nanyang Avenue, Singapore 639798, Singapore
}


\begin{abstract}
3D Dirac semimetals (DSMs) are promising materials for terahertz high harmonic generation (HHG). We show that 3D DSMs' high nonlinearity opens up a regime of nonlinear optics where extreme subwavelength current density features develop within nanoscale propagation distances of the driving field. Our results reveal orders-of-magnitude enhancement in HHG intensity with thicker 3D DSM films, and show that these subwavelength features fundamentally limit HHG enhancement beyond an optimal film thickness. This decrease in HHG intensity beyond the optimal thickness constitutes an effective propagation-induced dephasing. Our findings highlight the importance of propagation dynamics in nanofilms of extreme optical nonlinearity.
\end{abstract}

\pacs{Valid PACS appear here}
\maketitle
The terahertz (THz) regime has attracted much attention due to its broad range of potential applications, including electron acceleration~\cite{Zhang2020, Nanni2015, Sharma2018}, imaging~\cite{Cocker2013, Stantchev2020, Wu2019}, controlling ultrafast processes in materials~\cite{Kampfrath2011, Kampfrath2013, Zaks2012}, and next-generation communications~\cite{Nagatsuma2016, Zhang2016,  Yang2020}. These emerging technologies have motivated the quest to realize compact solid-state THz high-harmonic generation (HHG) platforms. The efficiency of HHG, which involves light emission at integer multiples of the input laser frequency, favors materials with strong optical nonlinearity. Three-dimensional Dirac semimetals (3D DSMs)~\cite{Liu2014a, Liu2014b, Borisenko2014, neupane2014, Zheng2016, Novak2015}, whose massless charge carriers result in extreme nonlinearity in the THz regime, is one such example. Recent experiments~\cite{Cd3As2THzHHG,Kovalev2019} and theory~\cite{APLphotonics_YeeSin, Zhang19, Lim_2020, Cheng2020} have verified the promise of 3D DSMs in realizing compact, highly efficient THz HHG sources. In particular, even for moderate driving fields of $\leq$10 MV/m, highly efficient generation of the 3rd~\cite{Cd3As2THzHHG} and up to the 7th~\cite{Kovalev2019} harmonic have been demonstrated using the 3D DSM Cd$_{3}$As$_{2}$, with a theoretical study predicting efficient HHG up to the 31st order~\cite{Lim_2020}. In contrast, conventional solid-state THz HHG platforms require driving fields $\gtrsim 1$ GV/m to generate comparable THz HHG intensities~\cite{Schubert2014, Hohenleutner2015}.

One arguable advantage of solid-state THz HHG sources is the ability to enhance output intensity by increasing propagation length through increased material thickness. The thickness of 3D DSMs can be tailored using well-known semiconductor film thickness control techniques~\cite{Meng18}.  However, this prospect of enhancing THz HHG in 3D DSMs remains unexplored.

Here, we show that orders-of-magnitude enhancements of the output THz HHG intensity from 3D DSMs is attainable by increasing the propagation length. Specifically, our results reveal that for a 10 MV/m driving field, the output intensities of the 3rd and 31st harmonics can be enhanced by factors of 144 and 28 respectively in going from a film thickness of 50 nm to 1500 nm.  Crucially, we show that highly nonlinear materials like 3D DSMs open up a regime of nonlinear optics in which extremely subwavelength features in the induced current density appear in nanoscale propagation distances -- an effect not seen in conventional nonlinear materials.  These subwavelength features manifest primarily as a phase-flip in the induced current density, resulting in an optimal thickness for HHG in 3D DSM nanofilms. Beyond this optimal thickness, output HHG falls rapidly due to destructive interference between the emitted waves. We show that this decrease in output HHG constitutes an effective propagation-induced dephasing mechanism, and that larger optimal thicknesses accompany stronger driving fields. Additionally, for fixed field strengths, the optimal film thickness for all harmonics falls within a narrow range of values, indicating that many harmonics can be simultaneously optimized through a single choice of film thickness.  Our findings highlight the importance of accounting for light propagation dynamics in highly nonlinear nanofilms in general -- not just 3D DSMs, and suggest that appropriate nanostructuring could yield still-larger THz HHG output intensities. Our work paves the way to efficient, solid-state THz light sources and optoelectronics based on 3D DSMs.

\begin{figure*}[ht!]
\centering
\includegraphics[width = 0.95\textwidth]{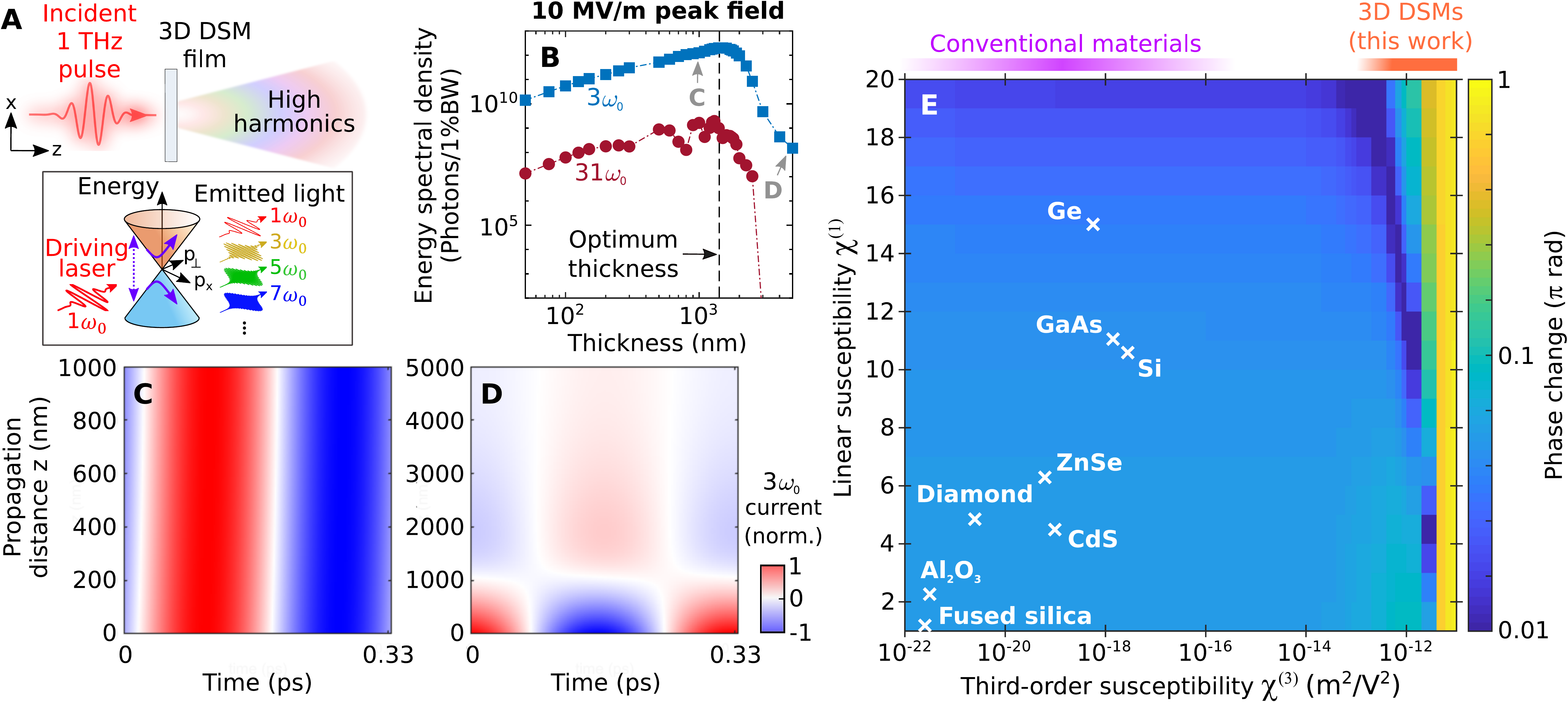}
\caption{Propagation-induced limits on HHG in 3D DSMs. (a) An $x$-polarized centered at 1 THz impinges on a 3D DSM thin film at normal incidence, resulting in the radiation of higher harmonics. In momentum space ((a) inset), the driving field induces intraband carrier oscillations (solid purple arrows) within and interband carrier transitions (dotted purple arrow) between the valence and conduction bands of the Dirac cone, resulting in the emission of light peaked at odd-integer multiples of the input frequency. (b) shows orders-of-magnitude enhancements in output intensity with increasing film thickness up to an optimal value of about 1500 nm, beyond which the harmonic output drastically falls. The existence of an optimal thickness arises due to the propagation-induced phase shift of the harmonic current (normalized 3rd harmonic current plotted in (c),(d) for 1 $\mu$m and $5~\mu$m) across the film thickness. In thinner films (c), the phase shift in the current as a function of propagation distance $z$ is insignificant. Thus, the emitted waves from different $z$ are in-phase and add constructively.  For films much thicker than the optimal thickness (d), a $\pi$-phase-flip occurs in the current density. The emitted radiation from opposite sides of this phase flip destructively interfere, resulting in the drastic decrease in HHG output with increasing film thickness seen in (b).  We find that the extreme subwavelength phase-flip of the induced current occurs within a regime of nonlinear optics accessible only by highly nonlinear materials like 3D DSMs; no such phase-flip occurs at \emph{extreme subwavelength propagation distances} in conventional nonlinear materials, which have weaker nonlinearity. We show this in (e), which plots the phase shift in the 3rd harmonic current for a 2500 nm-thick, nondispersive dielectric film as a function of its linear and third-order susceptibilities, $\chi^{(1)}$ and $\chi^{(3)}$ respectively, driven by an external field of 10 MV/m.  For conventional materials, we use the values of $\chi^{(1)}$ and $\chi^{(3)}$ found in Ref.~\cite{Boyd}.}
\label{fig_1}
\end{figure*}

\begin{figure*}[ht!]
\centering
\includegraphics[width =0.93 \textwidth]{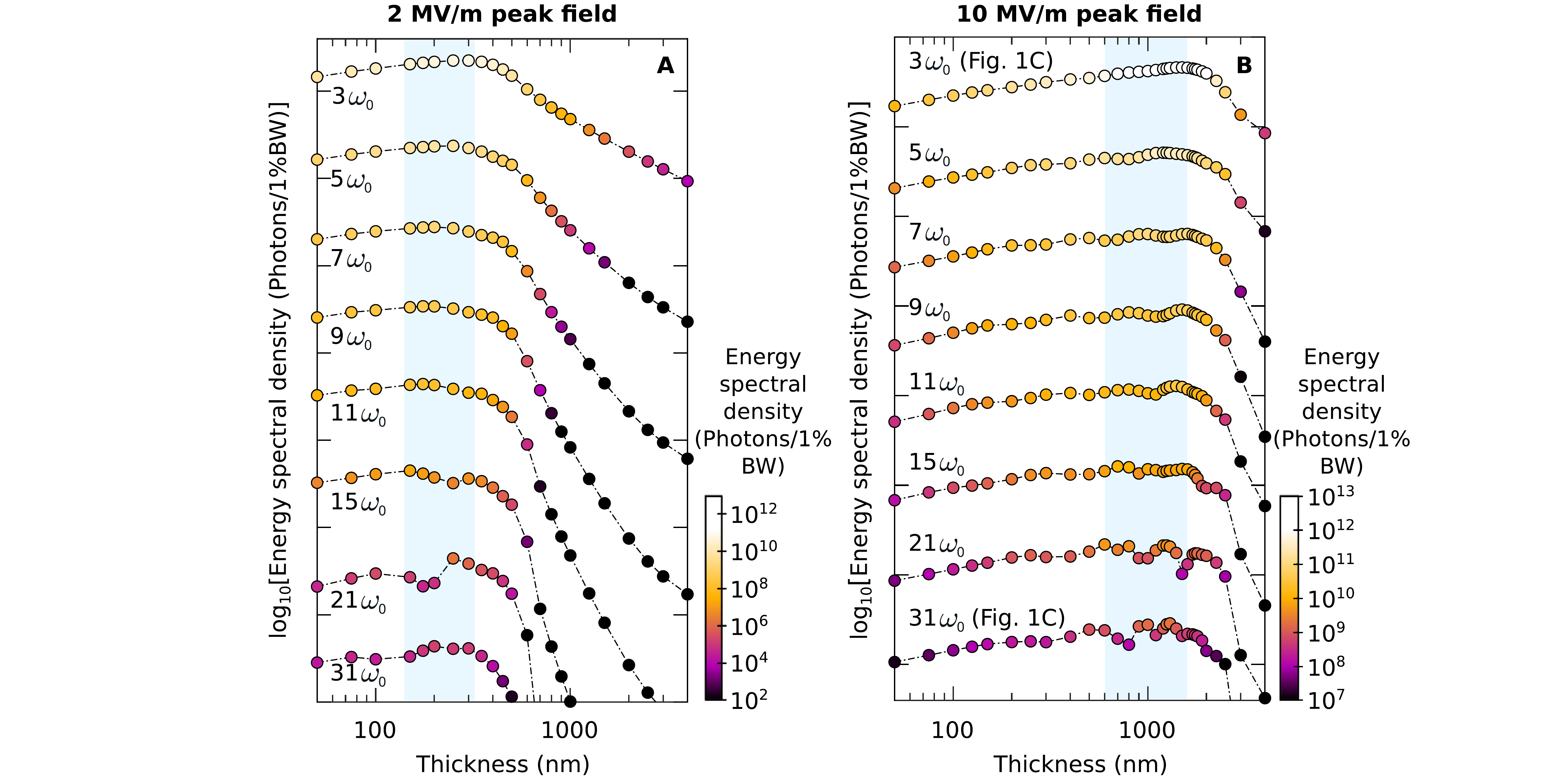}
\caption{Optimizing film thickness for HHG in 3D DSMs. (a) For a 3D DSM film driven by a pulse of peak amplitude 2 MV/m, the optimal thickness for all harmonics fall between 150 nm and 300 nm (blue shaded region). For a larger driving field of 10 MV/m (b), the optimal thickness of all harmonics lies between 600 nm and 1600 nm. The optimal thickness generally shifts to larger values when driven by stronger fields. In both panels, the $N$th curve from the top has been translated downwards by $10^{4(N-1)}$ for visual clarity; the color of the data points represent their actual magnitudes. We consider the same parameters as Fig.~\ref{fig_1}.}
\label{fig_2}
\end{figure*}

When an $x$-polarized THz pulse impinges upon a 3D DSM film at normal incidence (Fig.~\ref{fig_1}(a)), nonlinear currents are induced in the thin film, resulting in the emission of high harmonics. In momentum space (Fig.~\ref{fig_1}(a) inset), the driving laser field induces carrier oscillations within and transitions between the conduction and valence bands, giving rise to intraband and interband currents respectively, which emit light peaked at integer multiples of the input frequency. 

We model the interaction between the incident THz pulse and the 3D DSM film by solving Maxwell's equations and determine the nonlinear current density $J_{x}(z,t)$ via nonperturbative quantum theory. To confirm that the propagation-induced effects we observe are not a consequence of temperature or scattering, we consider the low temperature limit ($T\rightarrow0$ K) and no carrier scattering ($\tau\rightarrow\infty$), although our model can fully account for these effects. Under these conditions, the current density is given by~\cite{Lim_2020}:
\begin{widetext}
\begin{equation}
\begin{split}
J_{x}(z,t) =& -\frac{ge^{2}v_{x}}{6\pi^{2}\hbar^{3}v_{y}v_{z}}a_{x}(z,t)\Bigg{[} \ef^{2} -  \frac{e^{2}}{5}v_{x}^{2}a_{x}^{2}(z,t)\Bigg{]},~~~\qquad ev_{x}|a_{x}(z,t)| \leq \ef\\
J_{x}(z,t) =& -\frac{ge\ef^{3}}{6\pi^{2}\hbar^{3}v_{y}v_{z}}\mathrm{sgn}\big{[}a_{x}(z,t)\big{]}\Bigg{[} 1 -   \frac{\ef^{2}}{5e^{2}v_{x}^{2}a_{x}^{2}(z,t)}\Bigg{]},\quad ev_{x}|a_{x}(z,t)| > \ef,
\end{split}
\label{eqn_closed_form_Jx}
\end{equation}
\end{widetext}
where we have considered only the intraband current, which dominates when $\hbar\omega\ll2\ef$ ($\hbar$ is the reduced Planck's constant, $\ef$ is the Fermi energy, $\omega$ is the driving angular frequency), which is the case we study. Additionally, experiments~\cite{Cd3As2THzHHG,Kovalev2019} have shown that the intraband nonlinearities are dominant for high doping values in the THz regime. Specifically, we consider an incident pulse centered at $\omega_{0}=2\pi\times(1$ THz) and a Fermi energy $\ef=60$ meV, which satisfy $\hbar\omega_{0}\ll2\ef$. In Eq. (\ref{eqn_closed_form_Jx}), $g=4$ is the product of the spin and valley degeneracies, $e>0$ is the elementary charge, $v_{i}$ is the 3D DSM’s Fermi velocity along the direction $i\in\{x,y,z\}$, and $a_{x}(z,t)=-e^{-t/\tau}\int^{t}_{-\infty}e^{t'/\tau}E_{x}(z,t')dt'$ is the modified vector potential~\cite{Marini2017}. When scattering is neglected ($\tau\rightarrow \infty$), which is the case for our simulations, $a_{x}(z,t)$ reduces to the standard definition of the vector potential: $A_{x}(z,t)=-\int^{t}_{-\infty}E_{x}(z,t')dt'$. We compute the HHG spectra as radiation from the current distribution $J_{x}(z,t)$. This produces the same results as computing the HHG spectra from the reflected and transmitted electric fields (Supplemental Material (SM) Section I). We consider the experimentally measured $\mathrm{Cd_{3}As_{2}}$ Fermi velocities: $(v_{x},v_{y},v_{z})=(1.28,1.30,0.33)\times10^{6}$ m/s~\cite{Liu2014a}. Our choice of $\ef=60$ meV is attainable through chemical doping, and the film thicknesses considered have been realized using molecular beam epitaxy~\cite{Cd3As2THzHHG, Kovalev2019, Neubauer2016, Yuan2017, Nishihaya2019a, Nishihayaeaar5668, Liang2017}.  Few-cycle THz pulses with peak field strengths similar to those we consider are readily accessible using compact sources~\cite{single_cycle_1THz, OR_organic_crystals, HuangOptLett2013, Fulop_mJ_THz, Fulop_THz_OR}. Unless otherwise specified, we use these parameters throughout our work. We emphasize that our method fully incorporates the dispersive induced refractive index through the current density given by Eq.~(\ref{eqn_closed_form_Jx}).

\begin{figure*}[ht!]
\centering
\includegraphics[width = 0.9\textwidth]{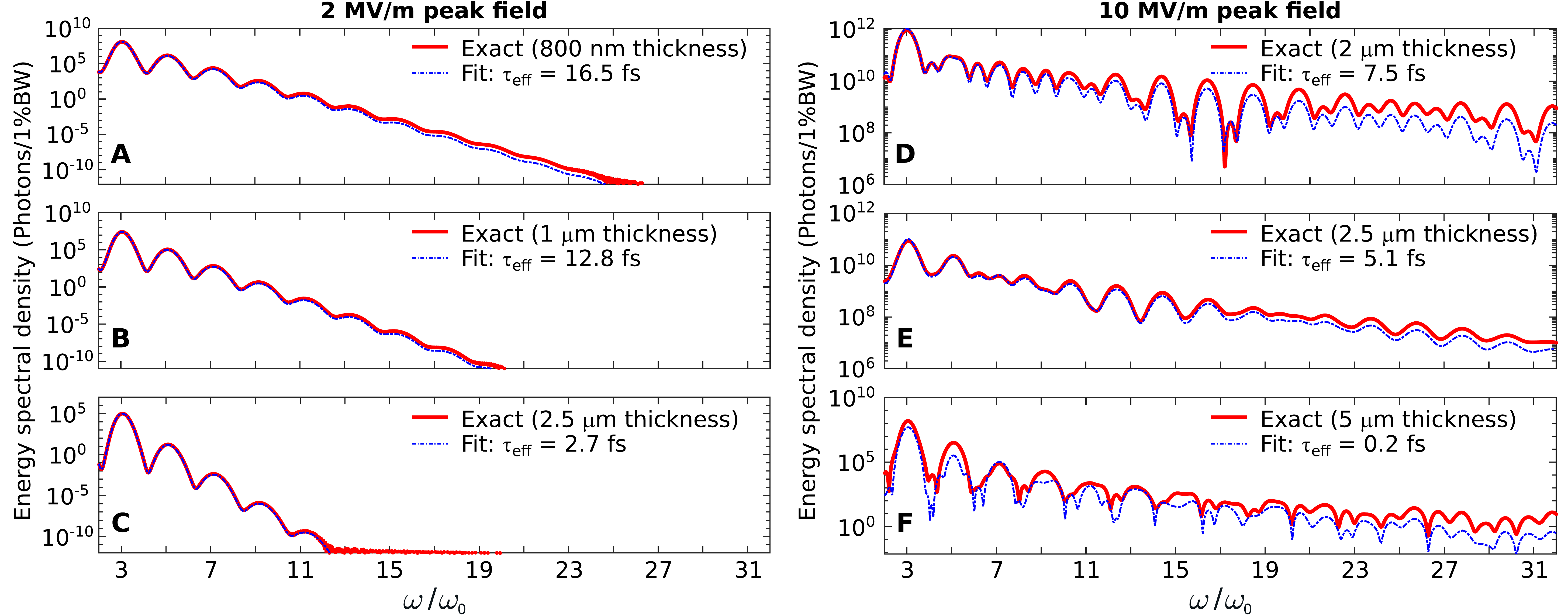}
\caption{Effective propagation-induced dephasing in THz HHG. For all panels, we plot the exact output HHG spectrum computed via Maxwell's equations as solid red curves, and the output HHG spectrum computed via Eq. (\ref{eqn_closed_form_Jx}) fitted to an effective scattering time $\tau=\teff$ -- while neglecting propagation effects -- as blue dashed-dotted curves. (a)-(c) show the spectra when 3D DSM films of thicknesses 0.8 $\mu$m, $1~\mu$m, and $2.5~\mu$m respectively are driven by a peak field of 2 MV/m. (d)-(f) show the spectra for film thicknesses of 2 $\mu$m, $2.5~\mu$m, and $5~\mu$m respectively, driven by a peak field of 10 MV/m. Our fitted values of $\tau_{\mathrm{eff}}$ are in good agreement with the exact spectra, implying that the reduced HHG can be captured by an effective dephasing time , and is thus due to propagation-induced dephasing effects. As expected, larger film thicknesses correspond to shorter $\tau_{\mathrm{eff}}$. We consider the same parameters as in Figs.~\ref{fig_1} and \ref{fig_2}.}
\label{fig_3}
\end{figure*}

Figure~\ref{fig_1}(b) shows our results for a $\mathrm{Cd_{3}As_{2}}$ film driven by an incident 2 ps-long (intensity full-width-at-half-maximum) 1 THz pulse of peak field 10 MV/m (in free space). In going from a film thickness of 50 nm to 1500 nm, we find that the 3rd and 31st harmonics are enhanced by factors of  144 and 28 respectively. However, beyond the optimal thickness of $\approx1.5~\mu$m, the output harmonic intensities rapidly diminish with increasing film thickness.

To investigate why the output harmonic intensities decrease beyond a certain film thickness, we plot the 3rd harmonic of the current density, defined as  $\mathrm{Re}[\tilde{J}_{x}(z,\omega)\exp(i\omega t)]$ for $\omega=3\omega_{0}$ in Figs.~\ref{fig_1}(c),~\ref{fig_1}(d) for film thicknesses of 1 $\mu$m and 5 $\mu$m respectively.  Here, $\tilde{J}_{x}(z,\omega)$ is the Fourier transform of $J_{x}(z,t)$. For a film thickness of 1 $\mu$m -- thinner than the optimal thickness -- we see from Fig. 1(c) that the current profile undergoes negligible phase shift with propagation distance.  Hence, at any given time, the radiation emitted by the current density at different propagation distances constructively interfere. However, for film thicknesses way beyond the optimal value (Fig.~\ref{fig_1}(d)), the current density undergoes a $\pi$-phase shift across the film. Consequently, the radiation emitted by currents at different $z$ destructively interfere, resulting in reduced output intensity. It is noteworthy that this phase-flip takes place at an extremely subwavelength propagation distance of 1 $\mu$m -- 100 times smaller than the driving wavelength -- and occurs within the skin depth of 1.4 $\mu$m (at 1 THz). 

Crucially, we find that this extreme subwavelength phase-flip only arises within a regime of nonlinear optics where highly nonlinear nanofilms, like 3D DSMs, are involved.  We show this in Fig.~\ref{fig_1}(e), where the phase shift of the 3rd harmonic current density in a 2500 nm-thick, nondispersive dielectric film driven by a peak field of 10 MV/m is plotted as a function of its linear and 3rd order susceptibilities, $\chi^{(1)}$ and $\chi^{(3)}$ respectively.  We see that a $\approx\pi$-phase flip over subwavelength propagation distances only manifests for materials with $\chi^{(3)}\gtrsim10^{-12}$ m$^2$/V$^2$, such as 3D DSMs.  In contrast, conventional materials, which possess relatively small 3rd-order susceptibility by comparison ($\chi^{(3)}\lesssim10^{-16}$ m$^2$/V$^2$), see negligible phase shifts over subwavelength thicknesses. See SM Section II for details of the simulations used to plot Fig.~\ref{fig_1}(e).

Figures~\ref{fig_2}(a) and~\ref{fig_2}(b) show the output HHG intensity as a function of film thickness for various harmonics. While the optimal thickness generally differs across harmonics, they lie within a relatively narrow range of values, implying that the output intensity of all harmonics can be simultaneously optimized with a single choice of film thickness. Considering a peak field of 2 MV/m (Fig.~\ref{fig_2}(a)), the maximum output intensity for each harmonic is achieved with film thicknesses ranging from 150 nm to 300 nm. Considering a peak field of 10 MV/m (Fig.~\ref{fig_2}(b)), the optimal film thickness lies between 600 nm and 1600 nm. The blue shading in Figs.~\ref{fig_2}(a) and~\ref{fig_2}(b) indicate these ranges of optimal thicknesses. Additionally, we observe that the optimal film thickness generally accompanies stronger driving fields.

Our investigations also reveal that the decrease in output HHG with increasing film thickness can be understood as a propagation-induced dephasing mechanism. We verify this by computing an effective scattering time $\teff$ in the absence of propagation. In Fig.~\ref{fig_3}, we show the excellent agreement between the exact spectrum obtained from Maxwell's Equations, and the spectrum obtained from Eq. (\ref{eqn_closed_form_Jx}) fitted to an effective scattering time $\tau=\teff$ while neglecting propagation. Considering an incident pulse with 2 MV/m peak field (Figs.~\ref{fig_3}(a)-\ref{fig_3}(c)), we find that the output spectra for film thicknesses of 800 nm, 1 $\mu$m, and 2.5 $\mu$m are effectively captured by $\teff=$ 16.5 fs, 12.8 fs, and 2.7 fs respectively. As expected, thicker films (beyond the optimal thickness) correspond to faster scattering times – a trend that holds for a peak field of 10 MV/m (Figs.~\ref{fig_3}(d)-\ref{fig_3}(f)): the spectra for film thicknesses 2 $\mu$m, 2.5 $\mu$m, and 5 $\mu$m are captured by $\teff=$ 7.5 fs, 5.1 fs, and 0.2 fs, respectively. Note that these values of $\teff$ are much shorter than the typical scattering times of $\approx150$ fs~\cite{Cd3As2THzHHG} in cleaner samples of $\mathrm{Cd_{3}As_{2}}$, indicating that propagation-induced dephasing is the dominant dephasing mechanism for thicker films. Neglecting propagation effects -- to show that the decrease in HHG output can be captured by $\teff$ -- involves assuming a driving field throughout the film that has the same field profile as the irradiated surface. Our procedure finds $\teff$ such that the area under the spectrum from $2\omega_{0}$ to $32\omega_{0}$ is the same as the area under the exact propagated numerical spectrum over the same frequency range.

Recent experiments~\cite{Cd3As2THzHHG,Kovalev2019} and theory~\cite{Lim_2020} have shown the bulk nature of 3D DSMs enables the output harmonic intensity to exceed that of 2D DSM graphene by over 10 times. Importantly, we find that increasing the thickness beyond previously considered values~\cite{Cd3As2THzHHG,Kovalev2019,Lim_2020} allows further enhancement of the HHG output intensity. However, our studies also reveal that using nanofilms of highly nonlinear materials, of which 3D DSMs are only one example, places us within a regime of nonlinear optics in which the induced current undergoes as much as a $\pi$-phase-shift over extreme subwavelength propagation distances. This phase-flip results in the existence of an optimal thickness, beyond which the HHG output intensity falls. We show that this decrease in HHG intensity can be captured by an effective propagation-induced dephasing time $\teff$. It is noteworthy that this propagation-induced dephasing occurs even over extremely subwavelength propagation lengths of $\sim100$ nm to $\sim1~\mu$m -- about 100 to 1000 times shorter than the central driving wavelength. In contrast, a recent theoretical study~\cite{Kilen2020} has shown that propagation-induced dephasing in conventional materials, which have weaker nonlinearity, only becomes significant when the film thickness far exceeds the central driving wavelength. We emphasize that their result is entirely consistent with our findings presented in this work: in conventional nonlinear materials, negligible phase-change of the induced current occurs over extreme subwavelength propagation distances, which in turn results in negligible propagation-induced dephasing on subwavelength scales. This explains why propagation-induced dephasing in conventional nonlinear materials is only expected for film thicknesses far exceeding the central driving wavelength.  Our study thus establishes the important role of light propagation effects on extreme subwavelength distances in this regime of nonlinear optics where highly nonlinear materials -- such as emerging topological semimetals beyond 3D DSMs~\cite{Yan2017, Li2017, Zhang2017, Huang2016,Li2017b, Li2018, Zhao2019, Zhang2018, Zhou2018, ChingHuaNodalHHG}, for instance -- are considered.

Our results suggest that restrictions on the optimal film thickness may be circumvented using nanostructured 3D DSM. By using a superlattice of 3D DSM nanofilms (e.g., 1D array of 3D DSM films) interleaved with other materials (e.g., metals or dielectrics), the output HHG intensity could be enhanced, even beyond the optimal film thickness in an unstructured 3D DSM film. Appropriately designed 3D DSM superlattices would also introduce additional versatility to the output HHG spectrum, for instance to selectively amplify or attenuate specific harmonics, enabling on-demand THz light-shaping on a chip-scale platform. 

In summary, we have shown that the output intensity of THz HHG in 3D DSMs can be enhanced by orders of magnitude with increased propagation length.  Crucially, using nanofilms made of highly nonlinear materials like 3D DSMs opens up a regime of nonlinear optics where extremely subwavelength features develop in the induced current density, an effect which is negligible in conventional nonlinear materials.  The phase-flip in current density results in the destructive interference of emitted radiation from different parts of the film. We verify that this phenomenon constitutes propagation-induced dephasing of the HHG process by computing an effective dephasing time $\teff$ for light generation in 3D DSMs. We also show that the optimal film thickness for all output harmonics lies within a relatively narrow range for a given driving field, and that this optimal film thickness generally increases for stronger fields. Our findings highlight the importance of including pulse propagation effects for highly nonlinear materials, including 3D DSMs and other bulk topological semimetals.  Furthermore, our findings suggest that appropriate nanostructuring could enable still-greater THz HHG output efficiency to be achieved.  Our work paves the way to efficient solid-state THz light sources and optoelectronics based on highly nonlinear material platforms like 3D DSMs.

\begin{acknowledgements}
L.J.W. acknowledges the support of the Agency for Science, Technology and Research (A*STAR) Advanced Manufacturing and Engineering Young Individual Research Grant (A1984c0043); and the Nanyang Assistant Professorship Start-up Grant.  J.L. and L.K.A. acknowledge funding from Singapore MOE Tier 2 Grant (2018-T2-1-007), MOE PhD RSS, and USA ONRG grant (N62909-19-1-2047). Y.S.A. acknowledges funding from SUTD Startup Research Grant (SRT3CI21163).
\end{acknowledgements}

\bibliographystyle{apsrev4-1} 
\bibliography{HHG_dephasing}  

\end{document}